\documentclass[twocolumn,superscriptaddress,prl]{revtex4-1}

\usepackage{graphicx}
\usepackage[usenames]{color}
\usepackage{multirow}
\usepackage{verbatim}
\usepackage{amsmath}
\usepackage{amsfonts}
\usepackage{amssymb}
\usepackage{mathrsfs}
\usepackage{url}
\usepackage{floatrow}
\usepackage{float}
\usepackage{setspace,ragged2e}

\begin{document}
\newcommand{\noct}{NiO$_6$}
\newcommand{\nii}{Ni$^{2+}$}
\newcommand{\niii}{Ni$^{3+}$}

\title{``Fraternal-twin'' ferroelectricity: competing polar states in hydrogen-doped samarium nickelate from first principles}
\author{Michele~Kotiuga}
\affiliation{Department of Physics and Astronomy, Rutgers, The State University of New Jersey, Piscataway, NJ, USA}
\author{Karin M. Rabe}
\affiliation{Department of Physics and Astronomy, Rutgers, The State University of New Jersey, Piscataway, NJ, USA}
\date{\today}
\begin{abstract}
  Reversible intercalation of hydrogen into  samarium nickelate, SmNiO$_3$ (SNO), has been of recent interest. Upon entering SNO, the hydrogen dissociates: the H$^+$ binds to an oxygen and the valence electron localizes on a nearby NiO$_6$ octahedron, resulting in a local dipole moment.
  In this work, we use first-principles calculations to explore the polar states of hydrogen-doped SNO at a concentration of 1/4 hydrogen per Ni. 
  The inherent tilt pattern of SNO and the presence of the interstitial hydrogen present an insurmountable energy barrier to switch 
  these polar states to their symmetry-related states under inversion. 
We find a sufficiently low barrier to move the localized electron to a neighboring NiO$_6$ octahedron,  a state unrelated by symmetry but equal in energy under epitaxial strain, resulting in a large change in polarization.
  We term this unconventional ferroelectric a ``fraternal-twin'' ferroelectric.  
\end{abstract}
\maketitle

In the prototypical perovskite oxide ferroelectrics, a soft polar mode gives rise to insulating polar states that can be switched by an applied electric field~\cite{Rabe2007}. In recent years, initially motivated by the search for ferromagnetic-ferroelectric multiferroics, there has been great interest in the design and discovery of ferroelectrics with alternative mechanisms. Examples include improper ferroelectrics, driven by a single nonpolar mode that couples to the polar mode at higher order (YMnO$_3$~\cite{Fennie2005}), hybrid improper ferroelectrics, involving the combination of multiple non-polar distortions (Ca$_3$Mn$_2$O$_7$~\cite{Benedek2011}); and charge ordering of multiple valence states~\cite{Efremov2004}, such as the iron in magnetite~\cite{Alexe2009}. Superlattice ordering has been explored as a means to break selected symmetries  to activate relevant nonlinear couplings or to make a particular ordering polar~\cite{Young2016,Park2017}. In each case, multiple symmetry-equivalent insulating polar states at the same energy are separated by barriers surmountable 
by accessible electric fields.

Recent experimental work has shown that samarium nickelate, SmNiO$_3$ (SNO) can be electron-doped by the reversible intercalation of hydrogen or lithium. Both experimental measurements and first-principles calculations show that the valence electrons of the intercalant localize on NiO$_6$ octahedra, changing the nominal valence of the Ni from 3+ to 2+, with the system remaining insulating for intercalant concentrations up to one per Ni~\cite{Shi2014, Zhang2018, Sun2018,Liu2019,Kotiuga2019}.  
First-principles calculations show that the favored sites for the intercalant ions are interstitial sites outside the \noct~octahedra, with a spatial separation between the localized electron and the positively charged intercalant~\cite{Zhang2018,Kotiuga2019}. Further, within the oxygen octahedron rotation pattern stabilized in the orthorhombic structure of undoped SNO, there are multiple choices of sites for the interstitial and corresponding choices for the NiO$_6$ octahedron to which the electron is added, yielding a large number of states comparable in total energy. 

  In this work, we systematically investigate low-energy states of hydrogen-doped SNO (H$_{1/4}$SmNiO$_3$)
  using first-principles calculations. The localization of the hydrogen valence electron on a nearby \noct~octahedron results in a local dipole pointing from the \noct~octahedron to the H$^+$ ion. These dipoles can be ordered to yield low-energy states with a net polarization. Due to the underlying 
  oxygen-octahedron tilt pattern that strongly stabilizes the orthorhombic structure of undoped SmNiO$_3$, a large energy barrier between symmetry-related polarization states prevents conventional ferroelectric switching. By applying cubic epitaxial strain to tune the potential energy landscape, we can equalize the energy of two states with roughly opposite polarization and the same octahedron-rotation pattern, switching the position of the 
  localized electron from one \noct~octahedron to a neighboring one. As the switching occurs between two states unrelated by symmetry, we refer to this as ``fraternal-twin ferroelectricity."

 \begin{figure*}[t]
\includegraphics[width=1\textwidth]{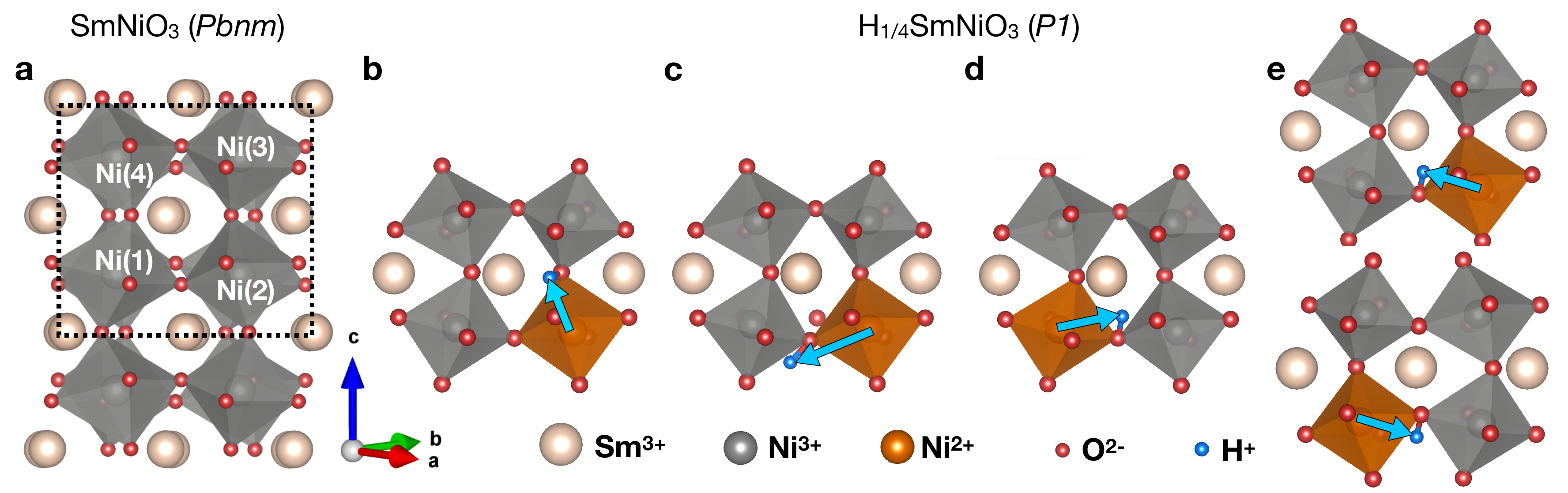}
\caption{(a) Bulk SmNiO$_3$ (SNO) with the space group $Pbnm$ (i.e. with an $a^-a^-c^+$ tilt pattern. The dashed box denotes the 20 atom unitcell with four nickel sites. (b)-(e) Four example structures of H$_{1/4}$SmNiO$_3$, i.e. one hydrogen added to the 20 atom unitcell, with the hydrogen bound to a basal or apical oxygen (relative to the $c$-axis). The hydrogen binds to an oxygen and its valence electron localizes on a nearby NiO$_6$ octahedron, colored orange. This results in a local dipole indicated by the cyan arrow.  (e) The two ``identical twin'' structures related by inversion.}
\label{fig:struct}
 \end{figure*}

Our first-principles DFT + $U$ calculations are carried out using the Perdew-Burke-Enzerhof (PBE) functional~\cite{Perdew1996,Perdew1997} as implemented in VASP~\cite{Kresse1996,Kresse1999} with an energy cutoff of 520 eV and the Sm\_3, Ni\_pv, O, and H projector augmented wave (PAW)~\cite{Blochl1994paw} potentials provided with the VASP package. The Sm\_3 PAW potential includes the $f$-electrons in the core. We include a Hubbard $U$ (within the rotationally invariant method of Liechtenstein et al.~\cite{Liechtenstein1995}) with $U$=4.6 eV and $J$=0.6 eV following our previous work~\cite{Kotiuga2019}. A 20-atom unit cell of SNO was used to accommodate the $a^-a^-c^+$ tilt pattern. We choose a G-type antiferromagnetic (AFM) ordering in our spin-polarized DFT calculations as our previous work has shown that the type of AFM ordering only effects the band width, but not the electron localization with electron doping. As SNO is paramagnetic in the temperature range of interest for ferroelectric properties, we do not study a ferromagnetic ordering. 
All structural relaxations were carried out with Gaussian smearing with $\sigma$ = 0.1 eV and a Monkhorst-Pack $k$-mesh of 6$\times$6$\times$4 such that the forces were less than 0.005eV/\AA, with lattice constants $a=5.278$\AA, $b=5.818$\AA, $c=7.421$\AA~ computed for undoped SNO. The density of states calculations were performed using  the tetrahedral method with Bl\"{o}chl corrections~\cite{Blochl1994}. The projected density of states (PDOS) plots were generated using a $\Gamma$-centered $k$-point mesh and the site projected scheme of pymatgen~\cite{Ong2013}. The polarization was calculated using the Berry phase method, as implemented in VASP. The branch choice was made such that the polarization aligned with the local dipole. 
The effects of epitaxial strain are studied with the strained-bulk method~\cite{Dieguez2005}, with in-place lattice parameters $a=b$ fixed while the $c$-axis is allowed to relax.

We begin by calculating the relative energetics of adding a hydrogen atom at various candidate interstital sites with corresponding choices for the location of the added electron. Specifically, in a $\sqrt{2}\times\sqrt{2}\times 2$ supercell of bulk SNO, we place one hydrogen atom in the plane bisecting the Ni-O-Ni bonds at the O for both apical and basal O sites, $\pm 1$\AA~away from the O along the crystallographic axes, and place the added electron in one of the two octahedra that include the O. We allow the internal coordinates to relax while fixing the lattice parameters.
The fixed lattice parameters allows us to compare the energetics of a number of structures and choose which ones to study further with epitaxial strain.
Epitaxial strain in the (001) plane is applied in the strained bulk method to tune the relative energies of the various configurations.



\begin{table}[!t]
\caption{Relative energetics of relaxed H$_{1/4}$SmNiO$_3$ constrained to the lattice parameters of bulk SmNiO$_3$, $a$ = 5.278\AA, $b$ = 5.818\AA, $c$ = 7.421\AA. The zero is set to be the low energy configuration, shown in Fig.~\ref{fig:struct}e.}
\centering
\setlength{\tabcolsep}{4pt}\def\arraystretch{1.1}
\begin{tabular}{c c c l c}
  \hline\hline
  H$^+$ between& Fig.&Ni$^{2+}$ &  \multicolumn{1}{c}{Polarization}&Energy \\ 
   &&(site)& \multicolumn{1}{c}{($\mu$C/cm$^2$)} &(meV/Ni)\\[.08cm]
  \hline

  Ni(2)\&Ni(3)&\ref{fig:struct}b&2&34 = \big|24,-17,17\big|&55\\[.08cm] 
  Ni(1)\&Ni(2) &\ref{fig:struct}c&2&31 = \big|-10,-12,-26\big|&26\\[.08cm] 
 
   Ni(1)\&Ni(2)&\ref{fig:struct}d&1&28 = \big|6,6,26\big|&24\\[.08cm] 

 Ni(1)\&Ni(2)&\ref{fig:struct}e&2&30 = \big|-7,-14,25\big|&0\\[.08cm] 
  \hline
\end{tabular}
\label{Table1}
\end{table}

Fig.~\ref{fig:struct} shows the crystal structure of bulk SNO, Fig.~\ref{fig:struct}a, and the four of the lowest-energy structures identified for H$_{1/4}$SNO, (Figs.~\ref{fig:struct}b-e and Table I), all of which are polar (see SI for more structures). If the octahedron-rotation distortions were not present, these four configurations would be symmetry equivalent, with a local dipole pointing from the electron on the \noct~to the adjacent H$^+$ interstitial. 
The polarizations and relative energetics of the structures pictured in Figs.~\ref{fig:struct}b-e are summarized in Table 1. The polarization primarily arises from the separation of the hydrogen ion and its valence electron. As this separation is governed by the chemical bond formed between the hydrogen and nearby oxygen as well as the Ni-O hybridization, the values for the different structures are quite similar, though not identical. The differences in energy stem from the underlying tilt pattern of SNO. 
The hydrogen can be placed between two NiO$_6$ octahedra with an in-phase tilt along the $c-$axis (Fig.~\ref{fig:struct}b) or an out-of-phase tilt along the $a-$ or $b-$axes (Figs.~\ref{fig:struct}c-e). Placing the hydrogen between two NiO$_6$ octahedra with a relative out-of-phase tilt results in a lower energy.
Furthermore, the in-phase tilts along the $c-$axis introduce a canting such that the  energy is lowest when the hydrogen is placed above the basal oxygen (Fig.~\ref{fig:struct}e, upper).  
Finally, the orthorhombicity of the cell results in different energies for the two structures shown in Fig.~\ref{fig:struct}d\&e,
which are related by an electron hop from Ni(1) to Ni(2) with a similar position of the H$^+$.

 \begin{figure}[t]
\includegraphics[width=\textwidth]{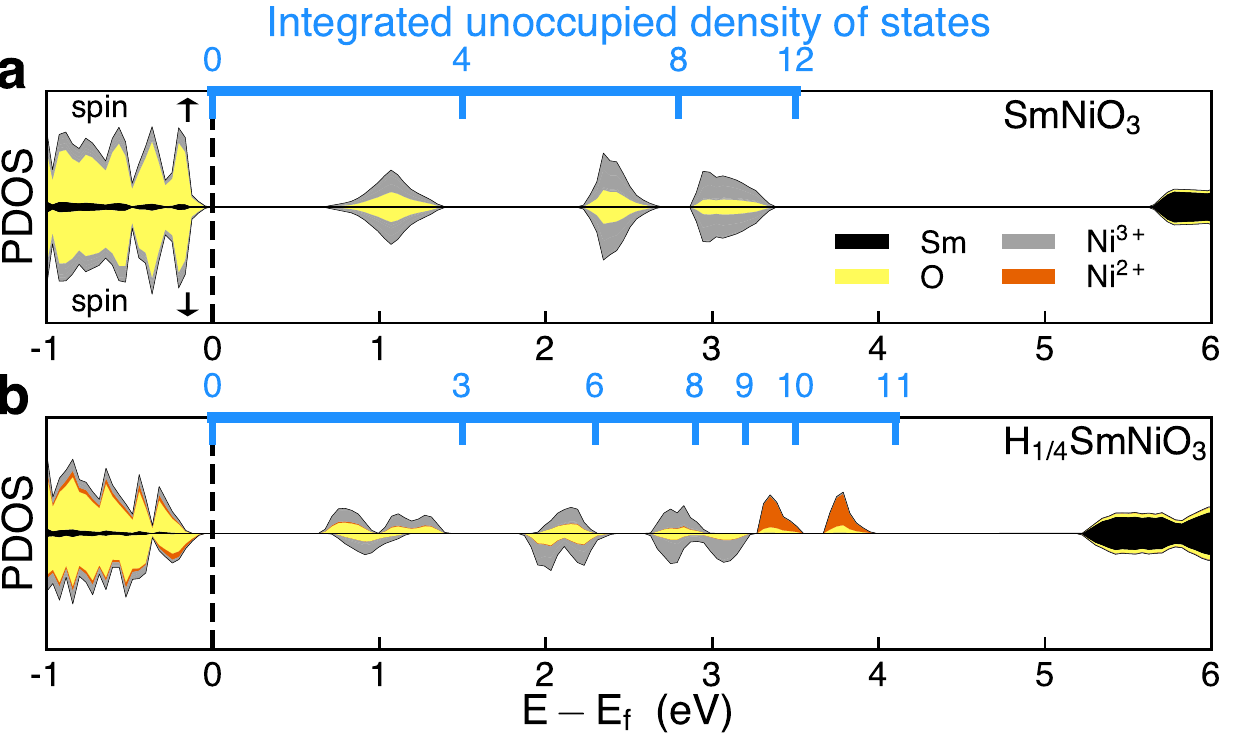}
\caption{Spin-polarized PDOS for (a) SNO and (b) H$_{1/4}$SNO (Fig.~\ref{fig:struct}e) for an AFM G-type magnetic ordering. The integrated unoccupied DOS is displayed along the top axis for $E-E_f\in[0,4]$. The element specific PDOS is shown in black for Sm, yellow for O, gray for Ni$^{3+}$ and orange for Ni$^{2+}$. }
\label{fig:dos}
 \end{figure}

For the lowest energy configuration, shown in Fig.~\ref{fig:struct}e, we also show the variant generated by inversion symmetry.
We can see that switching between these two structures would involve either an electron hop from one \noct~to another accompanied by a complete reversal of the octahedron-rotation pattern, or a large-scale displacement of the interstitial H$^+$ ion from one side of the octahedron to the other, see SI for energy barrier.

We find that these hydrogen-doped states are all insulating, with similar electronic structures. In Fig.~\ref{fig:dos}b, we show the spin-polarized DOS of the lowest energy state (Fig.~\ref{fig:struct}e). As in the spin-polarized DOS of undoped SNO, shown as a reference in Fig.~\ref{fig:dos}a, the computed gap is less than 1 eV. In the vicinity of the Fermi energy, the states are dominated by O-$2p$ and Ni-$3d$. In particular, the unoccupied states are Ni $e_g$ states hybridized with O-$2p$ states. 
In the undoped case, there are three unoccupied states per nickel. Each nickel has a magnetic moment close to 1$\mu_B$, suggesting low-spin Ni$^{3+}$. Due to the strong hybridization with the O-$2p$ states, we associate these states with the entire NiO$_6$ octahedron and to write the electronic configuration as $d^8\underline{L}$, i.e. $d^8$ with an oxygen ligand-hole. The added hydrogen results in the decrease in the number of unoccupied states from 4 to 3 in the feature closest to the Fermi energy.  The spin-down Ni-$3d$/O-$2p$ state associated with Ni(2) is moved to the top of the valence band and, dominated by oxygen, essentially fills the oxygen ligand-hole. The remaining unoccupied state associated with Ni(2), shown in orange, is pushed up in energy. The splitting of the lobes comes from the lowering of symmetry in these structures; however, as the electronic structure of the Ni$^{3+}$ sites is nearly unchanged, the change in the band gap relative to the undoped case is negligible.

 \begin{figure}[t!]
\includegraphics[width=\textwidth]{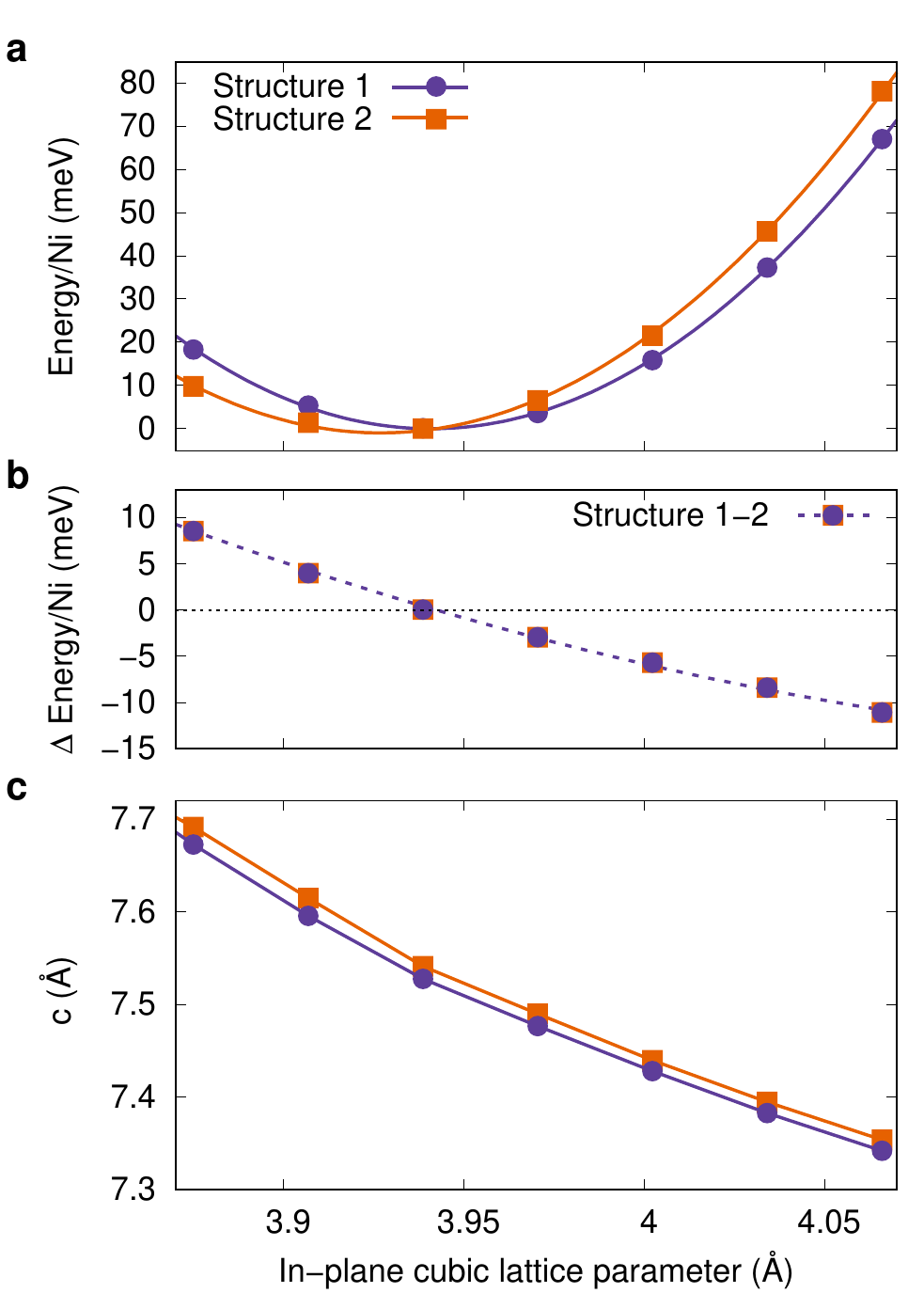}
\caption{(a) Total energy per nickel site as a function of the in-plane cubic lattice parameter for H$_{1/4}$SNO with the hydrogen valence electron localized on the octahedra containing Ni(1) (purple circles), like Fig.~\ref{fig:struct}d), and Ni(2) (orange squares), like Fig.~\ref{fig:struct}e, with quadratic fits. The zero is set to the lowest energy configuration. 
(b) The energy difference per nickel site for the electron localized on the octahedron containing Ni(1) or Ni(2) as a function of the in-plane lattice parameter. The solid line is the difference in the quadratic fits from panel (a). (c) The relaxed $c$-lattice parameter as a function of the in-plane lattice parameter for the electron localized on the octahedra Ni(1) \& (2). The lines serve only as a guide for the eye.}
\label{fig:EvA}
\end{figure}


For ferroelectricity, we require two states with different polarization at the same energy with a low barrier for switching. Here, the state related to the ground state by an electron hop, without other major structural rearrangements, is 24 meV per Ni higher in energy. Application of an epitaxial strain constraint offers the possibility of tuning the relative energy of these two states. In Fig.~\ref{fig:EvA}, we show the effect of cubic epitaxial strain on the structures of Fig.~\ref{fig:struct}d\&e, which we refer to as structure 1 and 2, respectively.
Fig.~\ref{fig:EvA}a shows that the total energy curves as a function of in-plane lattice parameter. The total energy curves of these two structures cross near their respective minima at 
$a=3.94$\AA, the calculated lattice constant for SrTiO$_3$ with DFT-PBE. This crossing can be more clearly seen in Fig.~\ref{fig:EvA}b, which shows the difference of these two energy curves. Due to the epitaxial constraint, this crossing occurs at an energy 70 meV/Ni higher than structure 2 with the bulk lattice parameters of SNO. Moreover for all values of $a$, the relaxed $c$-axis for the two structures is very close (within 0.02\AA), as shown in Fig.~\ref{fig:EvA}c. We note that 
at the bulk lattice parameters of SNO and $a>3.86$\AA~the OH bond is parallel to the $c$-axis, but for  $a<3.86$\AA~the low energy position of the hydrogen atom changes and the OH bond 
lies in the $ab$-plane. 
For clarity, the latter structures are reported in the SI and not included in Fig~\ref{fig:EvA}.  

\begin{figure}[t!]
\includegraphics[width=\textwidth]{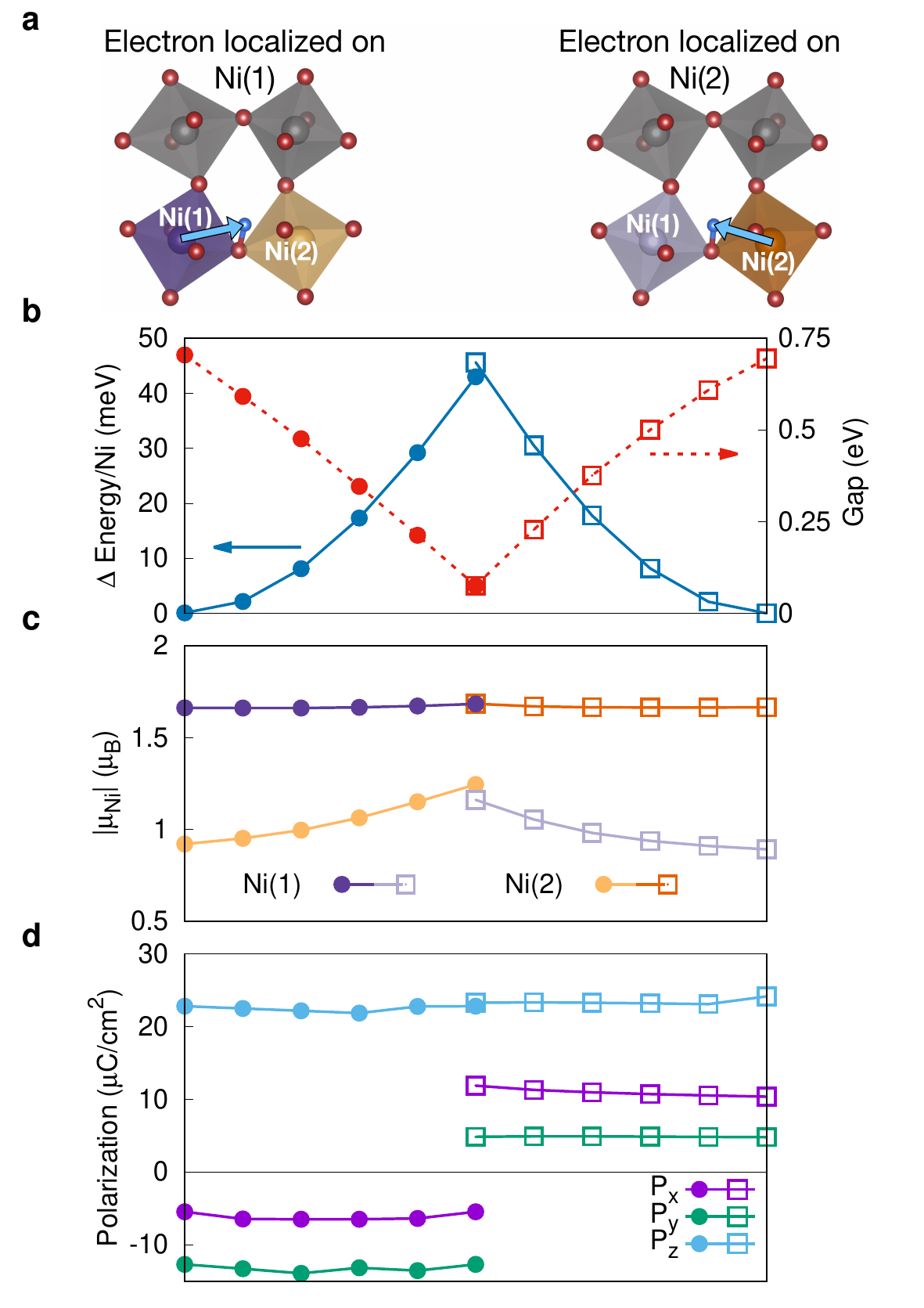}
\caption{(a) Unit cell of structures 1 (left) and 2 (right) omitting the Sm atoms. The closed symbols signal when the electron is localized on the octahedron containing Ni(1) and the open symbols  on the one containing Ni(2). (b) Change in total energy along the linearly interpolated switching path. (c) Magnetic moment of Ni(1), purple, and Ni(2), orange, along the switching path. The darker hues indicate a large magnetic moment (Ni$^{2+}$), whereas the lighter hues a smaller magnetic moment (Ni$^{3+}$). (c) The polarization along each Cartesian component through the switching path. }
\label{fig:switch}
\end{figure}

To characterize the switching, we 
linearly interpolate  between the structures 1 and 2 with $a=b=3.94$\AA~shown in Fig.~\ref{fig:switch}a. The position of the H$^+$ ion is allowed to relax so that it remains at an energetically favorable site,  while the rest of the cell is kept fixed with $a=b=3.94$\AA~and $c=7.52$\AA. 
Results for the energy of states along the path, shown in Fig.~\ref{fig:switch}b, give an energy barrier of $\sim$50 meV/Ni, corresponding to a coercive field on the order of 300 kV/cm. 
Through this switching path, the system remains insulating, with the band gap shown in Fig.~\ref{fig:switch}b. We can clearly track the position of the localized added electron through the Ni magnetic moments. In Fig.~\ref{fig:switch}c we observe a discrete jump, signifying that the electron hops from one NiO$_6$ octahedron to the other. At the midpoint of this path, the added electron can localize either on the octahedron containing Ni(1) or the one containing Ni(2), with a small energy difference due to the fact that the two states are not symmetry related. 
Finally, in Fig.~\ref{fig:switch}d, we track the polarization in each Cartesian component through this switching path. $P_z$ remains constant because the valence electron is hopping in the $ab$-plane, but we see a reversal of the polarization in the $ab$-plane, resulting in a sizable change in the polarization, $\Delta P=27\mu$C/cm$^2$.   

This energy barrier in this system is much lower than is typical of a charge-ordered ferroelectric, where strong coupling between the structure and the charge ordering tends to lock the hopping charge into a particular site~\cite{Qi2020}.  In this system, the Ni$^{3+}$O$_6$ octahedra surrounding the Ni$^{2+}$O$_6$ octahedron reduce the degree of structural relaxation, and the strong  hybridization between the Ni and O states aids the electron hop, resulting in a lower lattice coupling and barrier.

 Previous experimental work on electron-doped SNO suggests that this ferroelectric switching could be realized in the laboratory. Hysteresis at low biases has been observed in SNO on silicon~\cite{Hyeon2013}. It has been attributed to oxygen vacancies present in the system which effectively electron dope the system. 
Growing epitaxially-strained (001) SNO on SrTiO$_3$ applies the necessary strain to tune symmetry-inequivalent minima with very different polarizations to the same energy. The key will be to achieve a nearly homogeneous doping concentration throughout the thin film. 
 As the intercalant mobility is dependent on temperature~\cite{Zhou2016,Sun2018}, we expect the change in polarization to be more robust at lower temperatures.  
 Furthermore, the magnitude of the field should be chosen with care as it has been shown that in oxygen-deficient NdNiO$_3$ (NNO), the oxygen vacancies can be concentrated using an applied field~\cite{Kotiuga2019b} and in hydrogen-doped NNO and SNO on fluorine-doped tin oxide and indium tin oxide the H$^+$ distribution can be modulated with high speed electrical pulses~\cite{Zhang2020}.

   As the two structures are not symmetry related, 
 we refer to this type of ferroelectricity by the term ``fraternal-twin."
 The potential for fraternal-twin ferroelectricity in hydrogen-doped SNO arises from the rich configuration space of structures generated by the many inequivalent sites for the interstitial H$^+$ ion and its localized valence electron. In general, a polar structure does not ensure ferroelectricity as characteristics such as the tilt pattern or charge ordering can result in a insurmountable  barrier to switching to the symmetry-related state. 
  Fraternal-twin ferroelectricity provides an alternative mechanism for change in polarization with an electric field in materials when conventional ferroelectricity is thwarted by high energy barriers. Materials with a number of local minima close in energy,  for example ionic conductors, are ideal candidates to observe this phenomenon. In fact, past work has observed ferroelectric transitions in ionic conductors~\cite{Scott1980,Baranov1989,Stefanovich1993,Zhou2020}.  

We have demonstrated the possibility of polarization switching in epitaxially strained H$_{1/4}$SNO between two symmetry-inequivalent states, which we term ``fraternal-twin ferroelectricity." Specifically, the change in polarization derives from the added valence electron hopping from one NiO$_6$ to another, with the two structures not related by symmetry due to the underlying oxygen-octahedron rotation pattern. 
In the age of materials design, these results pave the way for a new class of designed ferroelectrics.

\begin{acknowledgments}
We thank Shriram Ramanathan for useful discussions and acknowledge financial support from Office of Naval Research Grant N00014-17-1-2770.
\end{acknowledgments}

\bibliographystyle{apsrev4-1}
\bibliography{FE_HSNO}

\clearpage

\onecolumngrid

\appendix

\centering{\bf Supplemental Materials:``Fraternal-twin'' ferroelectricity: competing polar states in hydrogen-doped samarium nickelate from first principles}
\flushleft

\section{Interstitial sites}
\begin{figure}[!h]
\center
\vspace*{-7mm}
\includegraphics[width=0.5\textwidth]{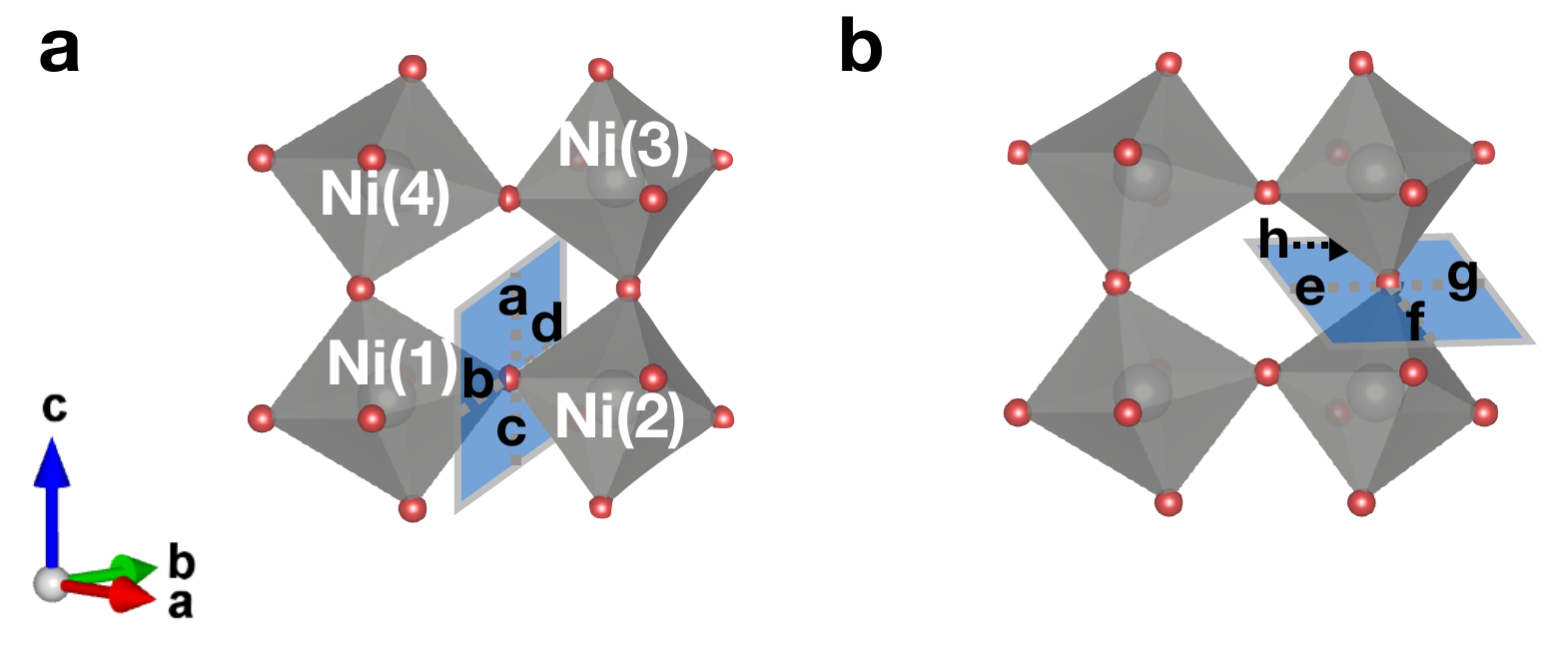}
\vspace*{-3mm}
\caption{We add a hydrogen atom at various candidate interstitial sites. Specifically, in a $\sqrt{2}\times\sqrt{2}\times 2$ supercell of bulk SNO, we place one hydrogen atom in the plane bisecting the Ni-O-Ni bonds at the O for both apical and basal O sites, $\pm 1$\AA~away from the O along the crystallographic axes, shown by the labels a-h}
\label{fig:sites}
\end{figure}
\vspace*{-6mm}
\begin{figure}[!h]
\center
\includegraphics[width=0.85\textwidth]{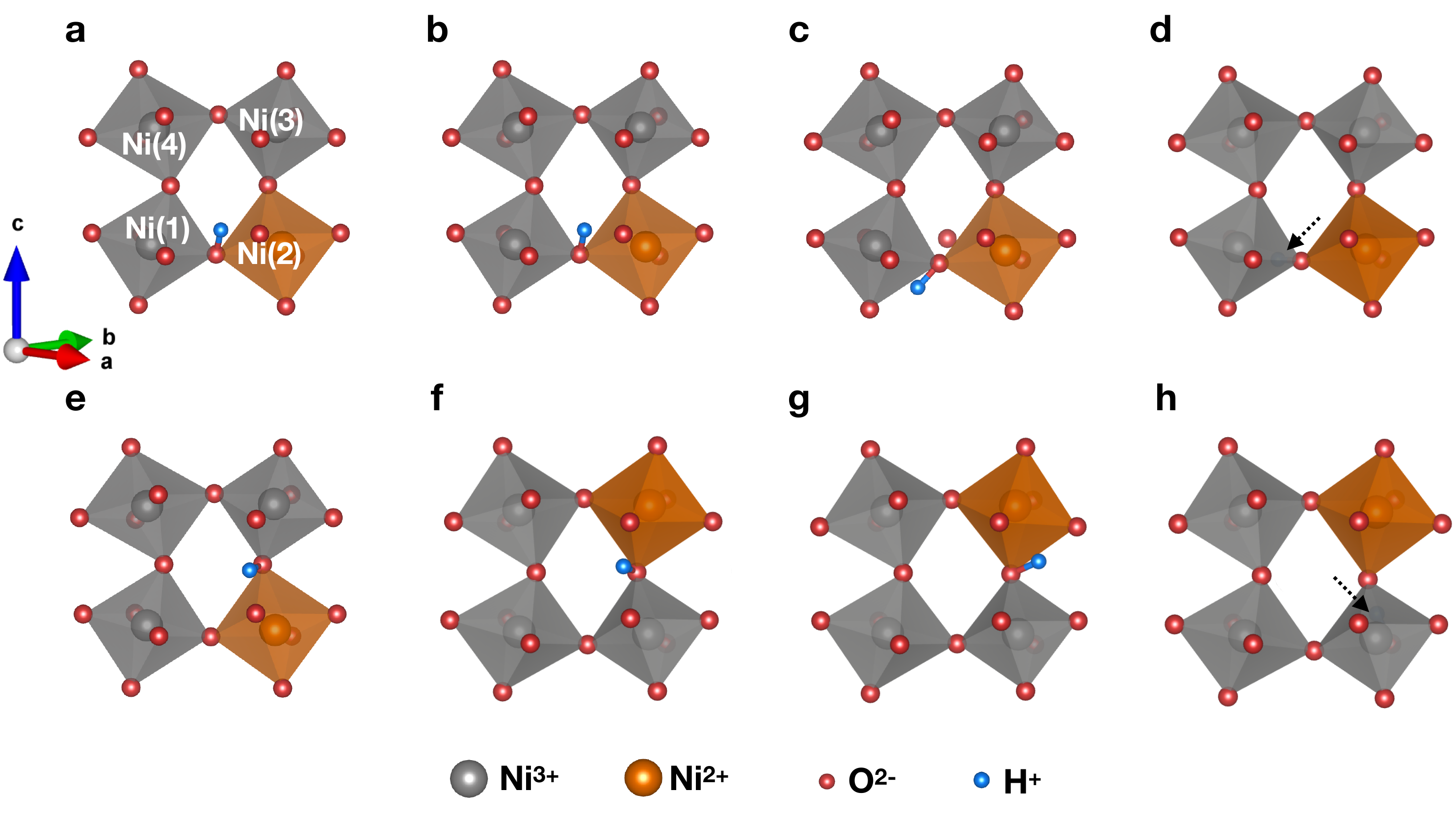}
\vspace*{-3mm}
\caption{Relaxed metastable H$_{1/4}$SmNiO$_3$ structures using the initial positions in fig.~\ref{fig:sites} without restricting on which NiO$_6$ octahedron the hydrogen valence electron localizes, for clarity we omit the Sm atoms. The fig.~\ref{fig:sites} labels correspond to the panels. Note: the initial starting positions (a) and (b) relax to the same structure and in panels (d) and (h) the hydrogen ion is behind the NiO$_6$ octahedron with an arrow pointing to its position. Panel (a) \& (b) are the same as Fig. 1e of the main text; panel (c) is the same as Fig. 1c of the main text; and, panel (e) is the same as Fig. 1b of the main text}
\label{fig:structSI}
\end{figure}

\newpage 

\begin{table}[!t]
\caption{Relative energetics of relaxed H$_{1/4}$SmNiO$_3$ constrained to the lattice parameters of bulk SmNiO$_3$, $a$ = 5.278\AA, $b$ = 5.818\AA, $c$ = 7.421\AA~without restriction the location of the localized hydrogen valence electron. The zero is set to be the low energy configuration,  Fig.~1a.}
\centering
\begin{tabular}{c c c c}
  \hline\hline
  H$^+$ starting postiton  &Ni$^{2+}$ &Energy \\ 
   Fig.~\ref{fig:sites}&(site) &(meV/Ni)\\
  \hline

  a&2&0\\
  b&2&0\\
  c&2&26\\
  d&2&46\\
  e&2&55\\
  f&3&55\\
  g&3&215\\
  h&3&76\\
   \hline
\end{tabular}
\label{Table1}
\end{table}

\begin{figure}[!h]
\center
\includegraphics[width=0.5\textwidth]{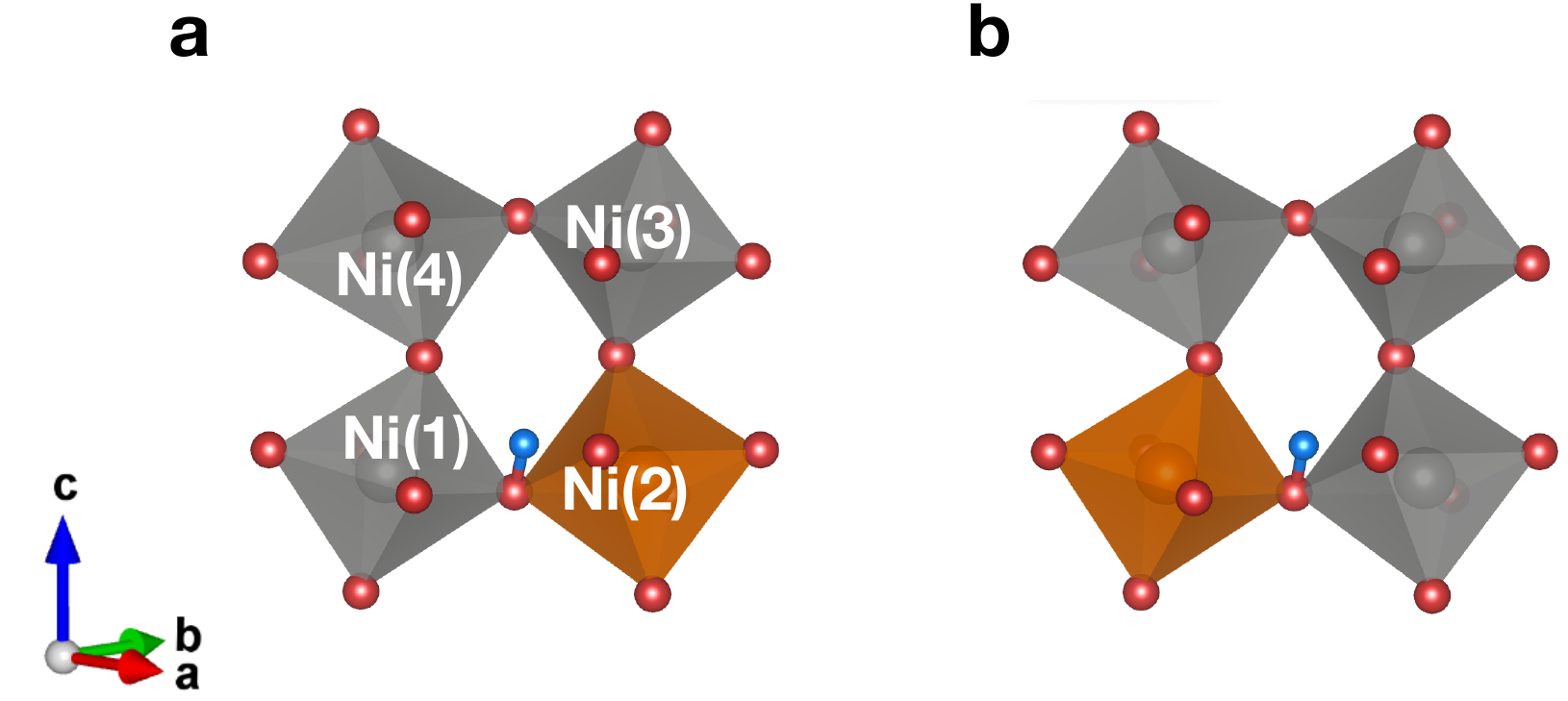}
\vspace*{-3mm}
\caption{(a) The same as Fig.~\ref{fig:structSI}a and Fig.~1e in the main text in which the electron is localized on the octahedron containing Ni(2). (b) The same as  Fig.~1d in the main text in which the electron is localized on the octahedron containing Ni(1), for clarity we omit the Sm atoms. }
\label{fig:e}
\end{figure}

Finally, we localize the hydrogen valence electron on the other NiO$_6$ octahedron neighboring the hydrogen ion, Fig.~\ref{fig:e}b. The resulting structure is 24 meV/Ni higher in energy than the lowest energy structure, Fig.~\ref{fig:e}a.

\clearpage
\section{Energy barrier to switch H$_{1/4}$SNO tilt pattern}

To calculate the energy barrier to switching the tilt pattern H$_{1/4}$SNO, we construct a path to interpolate between the structures Fig. 1e, upper and lower in the main text using the bulk lattice parameters of SNO. Linearly interpolating the posistions of the Sm, Ni and O atoms, we allow the position of the H$^+$ to relax so that the interstitial is in a low energy position. The hydrogen atom remains at an interstitial position that is between two octahedra canting toward one another. This site does not smoothly displace along this linearly interpolated path as such an interstitial position disappears at the midpoint of this linearly interpolated path and is a structure without any octahedral tilting. As the tilting switches the interstitial site jumps to the other side of the neighboring octahedra. To construct a smooth path, at this midpoint we freeze all the Sm, Ni and O atoms and move the H in an arc around the nearest O atom in the plane that contains the two low-energy interstitial sites and the O atom. We note that this path has no guarantee to be the lowest energy path connecting these two sites. See Fig.~\ref{fig:switchSI}a for structures illustrating this switching path.

\vspace{3mm}

We find an energy barrier larger than 1eV/Ni, a very larger energy barrier that cannot be switched with an applied electric field,Fig.~\ref{fig:switchSI}b. Furthermore, the system does not remain insulating along the switching path. The system becomes metallic as the added electron delocalizes.   This can be seen by inspecting the the magnetic moment of Ni(1) and Ni(2), Fig.~\ref{fig:switchSI}c. In the middle of the switching path the two moments are equal and there is no  discrete jump (as seen in Fig. 3c of the main text). This delocalization is related to the suppressing and removal of the octahedral tilts along this switching path. In the the rare-earth nickelates, it is known that the tilt angles are tied to the insulating states and that, at a given temperature, the metallic phases occur for smaller tilt angles.

\begin{figure}[!h]
  \center

\includegraphics[width=\textwidth]{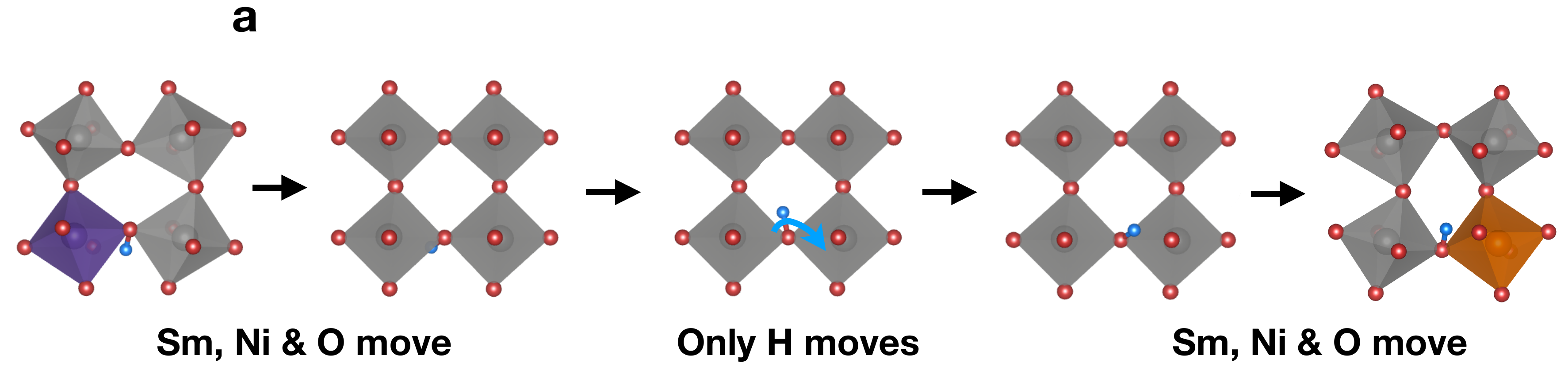}
\end{figure}
\begin{figure}[!h]
\center
  \vspace*{-11mm}
\includegraphics[width=0.76\textwidth]{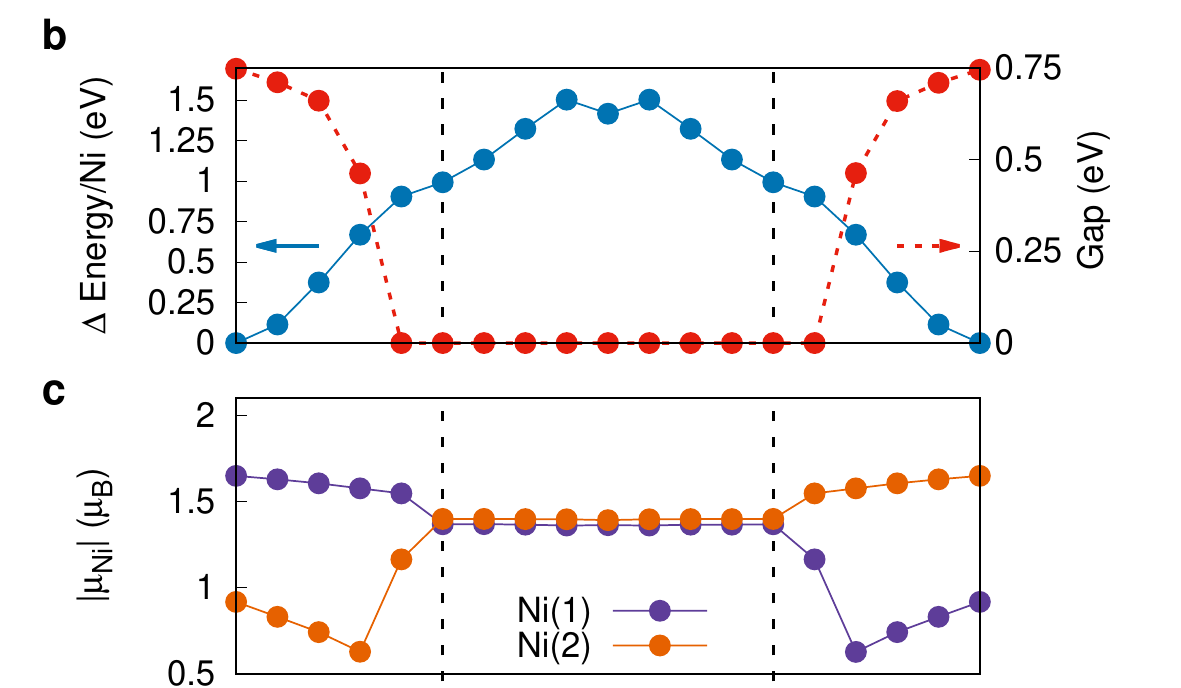}
\caption{(a) Structures illustrating the switching path between Fig. 1e, upper and lower in the main text, for clarity we omit the Sm atoms.. When all the NiO$_6$ octahedra are grey there is no electron localization. (b)   Change in total energy (blue) and gap (red) along this switching path. (c) The magnetic moment of Ni(1) and Ni(2) showing that the position of the added electron switches when the tilt pattern is inverted, but delocalizes in the process.}
\label{fig:switchSI}
\end{figure}

\clearpage
\section{Interstitial low energy site as a function of in-plane strain}
\begin{figure}[!h]
  \center
  \vspace*{-3mm}
\includegraphics[width=0.8\textwidth]{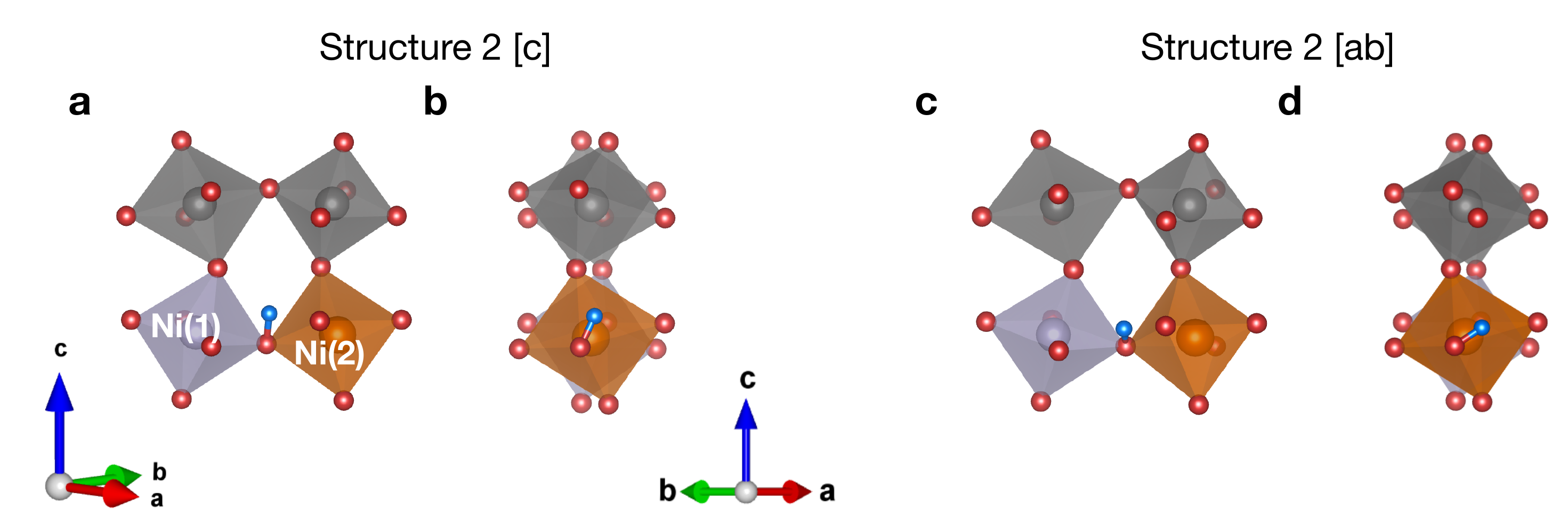}
\caption{Structure 2 [c] (a) front and (b) side views. Structure 2 [ab] (c) front and (d) side views. The dark orange octahedra signify the location of the added electron. Structure 1 [c] \& [ab] have the electron localized on Ni(1). For clarity, we omit the Sm atoms.}
\label{fig:ap_ba}
\end{figure}

For smaller in-plane lattice constants, we find that the position of the interstitial changes: the OH bond tilts down toward the $ab$-plane and away from the $c$-axis. Here, we add the total energy as a function of in-plane lattice parameter for this new minima to Fig. 3 of the main text. We call the structures from the main text ``structures 1\&2 [c]'' and these new structures  ``structures 1\&2 [ab],'' see Fig.~\ref{fig:ap_ba}. We see that when the OH bond cants toward the $ab$-plane, they have the same energy at $a=3.84$\AA. We do not explore fraternal-twin ferroelectric switching at this in-plane lattice parameter as it less stable than the structures at $a=3.94$\AA. The values of $a$ where points are missing from the plots indicate that we were not able to stabilize that structure at that in-plane lattice parameter.  

\begin{figure}
\vspace*{-5mm}
\floatbox[{\capbeside\thisfloatsetup{capbesideposition={right,center},capbesidewidth=6cm}}]{figure}[\FBwidth]
{\caption{(a) Total energy per nickel site as a function of the in-plane cubic lattice parameter for H$_{1/4}$SNO with the hydrogen valence electron localized on Ni(1) (structure 1 - purple circles)), and Ni(2) (structure 2 - orange squares), with quadratic fits. The zero is set to the lowest energy configuration. For each structure the OH bond can either be aligned with the $c-$axis [c], or in the $ab$-plane [ab]. As the in-plane lattice constant decreases, the [ab] structures become more favorable. (b) The energy difference per nickel site for the electron localized on the octahedron containing Ni(1) or Ni(2)  as a function of the in-plane lattice parameter. The differences are taken with respect to ``structure 2 [c]''. The dashed lines are the differences in the quadratic fits from panel (a). (c) The relaxed $c$-lattice parameter as a function of the in-plane lattice parameter for the electron localized on Ni(1) \& (2). The lines serve only as a guide for the eye. The [ab] structures have a larger relaxed $c$ lattice parameter. }}
{\includegraphics[width=0.48\textwidth]{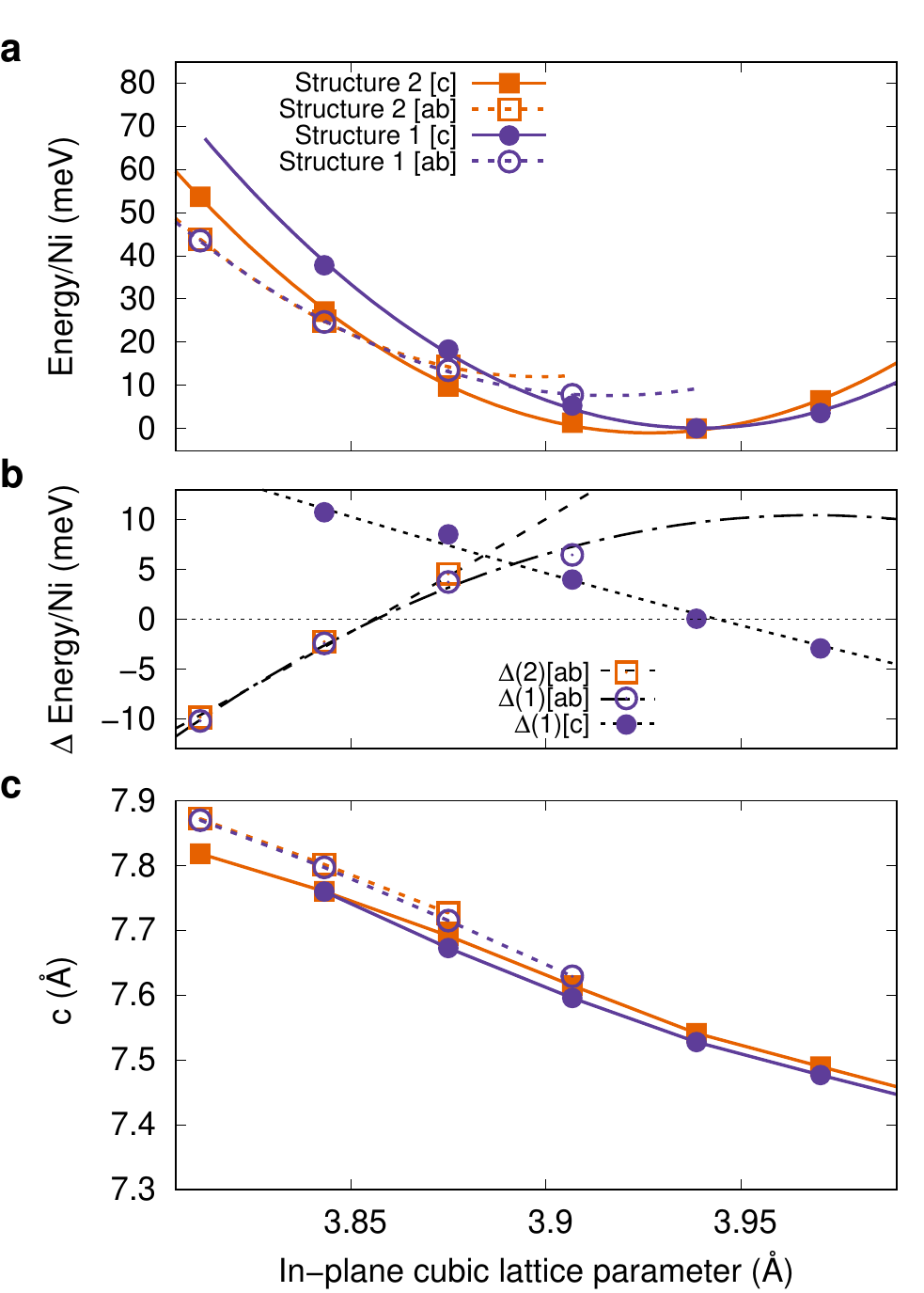}}
\label{fig:strain}
\end{figure}

\end{document}